\documentclass[12pt]{article}
\usepackage{graphicx}
\DeclareGraphicsExtensions{.pdf, .jpg, .png}
\usepackage{amssymb,amsmath}
\usepackage{epstopdf}
\usepackage{subfigure}
\usepackage{rotating}
\usepackage{color}
\usepackage{lscape,graphicx}
\usepackage{natbib}
\bibpunct{[}{]}{,}{n}{,}{,}

\title{Mechanical control of the directional stepping dynamics of the kinesin motor}
\begin{document}

\begin{center}
{\bf{\Large{Mechanical control of the directional stepping dynamics of the kinesin motor}}}
\end{center}

\vspace{.5in}

\begin{center}
\begin{tabular}[t]{c@{\extracolsep{2em}}c}
  Changbong Hyeon$^{1,2}$ and Jos{\'e} N. Onuchic$^1$\\
\end{tabular}
\smallskip
\begin{small}
\begin{tabular}{rl}
\parbox[t]{13cm}{
\begin{center}
$^1$  Center for Theoretical Biological Physics, University of California at San Diego, La Jolla, CA  92093\\
$^2$  Department of Chemistry, Chung-Ang University, Seoul 156-756, Republic of Korea
\end{center}
}\\
\end{tabular}
\end{small}
\bigskip
\end{center}

\baselineskip = 16pt

\begin{abstract}
Among the multiple steps constituting the kinesin's mechanochemical cycle, one of the most interesting events is observed when kinesins move an 8-nm step from one microtubule (MT)-binding site to another. 
The stepping motion that occurs within a relatively short time scale ($\sim 100$ $\mu s$) is, however, beyond the resolution of current experiments, therefore a basic understanding to the real-time dynamics within the 8-nm step is still lacking. 
For instance, the rate of power stroke (or conformational change), that leads to the undocked-to-docked transition of neck-linker, is not known, and the existence of a substep during the 8-nm step still remains a controversial issue in the kinesin community. 
By using explicit structures of the kinesin dimer and the MT consisting of 13 protofilaments (PFs), we study the stepping dynamics with varying rates of power stroke ($k_p$). We estimate that $k_p^{-1}\lesssim 20$ $\mu s$ to avoid a substep in an averaged time trace.  
For a slow power stroke with $k_p^{-1}>20$ $\mu s$, the averaged time trace shows a substep that implies the existence of a transient intermediate, which is reminiscent of a recent single molecule experiment at high resolution. We identify the intermediate as a conformation in which the tethered head is trapped in the sideway binding site of the neighboring PF. 
We also find a partial unfolding (cracking) of the binding motifs occurring at the transition state ensemble along the pathways prior to binding between the kinesin and MT. 
\end{abstract}

\section*{Introduction}

First recognized via their close relationship between ATPase activity and organelle transport along MTs \cite{Brady85Nature,Vale85Cell}, 
kinesins have received broad attention as a prototype of molecular motors 
for the last two decades.  
Recent single molecule (SM) experiments have shown that, with each stepping motion being strongly coupled to the ATP, the kinesin moves toward the (+)-ends of MTs by taking discrete 8-nm steps \cite{Visscher99Science,Block03Science,Cross05Nature,BlockPNAS06,Kaseda03NCB} in a hand-over-hand fashion \cite{Kaseda03NCB,Block03Science,Yildiz04Science}. 
Although the ultimate understanding to the kinesin's motility is still far from the completion, the SM experiments \cite{Visscher99Science,Block03Science,Block03PNAS,Kaseda03NCB,Cross05Nature,BlockPNAS06,Yildiz04Science} together with the series of kinetic ensemble measurements \cite{Gilbert94Biochem,Taylor95Biochem,Moyer98BC,Cross04TBS} and theoretical studies \cite{Prost97RMP,Fisher01PNAS,Fisher05PNAS,Gao06PNAS,Kolomeisky07ARPC,Hyeon07PNAS} begin providing glimpses to the physical principle of how kinesin walks.

Along the kinesin's mechanochemical cycle (SI Fig. 7), one of the main observations of the SM experiments is the stepping dynamics that enables the kinesin to move forward. 
The actual time spent for the stepping motion itself ($\lesssim 100$ $\mu s$), compared to the ATP binding and hydrolysis ($\gtrsim 10$ $ms$), is however too short to detect the details of the dynamics with the spatial and temporal resolution of current instruments. Thus, it is still difficult to answer many basic questions \cite{Block07BJ} related to the stepping dynamics. Some of those questions are : 
(a) During the 8-nm step, 
how does the swiveling motion of the tethered head occur? 
Does any detectable substep exist that reflects a transient intermediate? \cite{Nishiyama01NCB,Coppin96PNAS,Cross05Nature} 
(b) What fraction of the time and length scales is contributed from the power stroke and the diffusional search? 
(c) Does the kinesin walk parallel to the single PF or walk astride using two parallel PFs? \cite{Howard93JCB,Sheetz95BJ} 
To shed light on these questions,
we propose to take advantage of the native topology of kinesins and MTs. 
For instance, our previous study 
has clarified the regulation mechanism between the two heads \cite{Hancock99PNAS,BlockPNAS06} by using the native topology-based two-head bound model of kinesin on the MT \cite{Hyeon07PNAS}. 
Following the same line of thought, we show that even the dynamical pathways of kinesins reflect ``the topological constraints emanating from the molecular architecture.''
In the present work, 
we have adapted our previous two-head bound model \cite{Hyeon07PNAS} to study the stepping dynamics of kinesins on the 13-PF MT.  

According to the ``neck-linker docking model \cite{ValeNature99,Asenjo06NSMB,Tomishige06NSMB},'' the stepping motion of kinesin is initiated by the undocked-to-docked transition of the neck-linker at the MT-bound head (power stroke), and the rest of the binding process of the tethered head to the next MT-binding site is accomplished via a diffusional search. The whole stepping dynamics is typically interpreted as a combination between the directional motion of power stroke and the nondirectional diffusional search. 
At the molecular level, the length of the neck-linker shrinks gradually, with a progress of power stroke, making transition from a disordered to an ordered state; the resulting configurations bias the diffusional motion of the tethered head in one direction. 
Thus, we can rather interpret the power stroke as a time dependent boundary condition that biases the diffusional motion, considering that the tethered head diffuses on the time dependent energy landscape to search for the next MT-binding site.

To study the dynamics, we first obtain through simulations potentials of mean force (PMF) felt by the kinesin's tethered head on the MT at two extreme cases; one is the PMF with fully disordered neck-linker ($F_{\lambda=0}(x,y,z)$) and the other with ordered neck-linker ($F_{\lambda=1}(x,y,z)$), where $\lambda$ is a parameter that specifies the degree of the neck-linker being zippered. 
The kinesin structures with a disordered ($\lambda=0$) and an ordered neck-linker ($\lambda=1$) at the MT-bound head are generated by switching on and off the native contacts inside the green circles in Fig. 1B (see Results and Methods for more details). 
We subsequently perform Brownian dynamics simulations of a quasi-particle representing the tethered head on a PMF that is mixed between the two extreme cases. 
The disordered-to-ordered transitions are mimicked by mixing the two PMFs, $F_{\lambda=0}(x,y,z)$ and $F_{\lambda=1}(x,y,z)$, in a time-dependent manner.
We show that because of the multiple binding sites on the MT surface, the tethered head has a chance of misbinding to other MT-binding sites. 
To avoid such an intermediate, we argue that the rate of power stroke ($k_p$) should be faster than the sampling rate on the MT surface ($k_E$). 
Throughout the paper, we designate the rate of power stroke and the rate of space exploration by diffusion as $k_p$ and $k_E$, respectively.

The present work provides scenarios on how the dynamic pathways of the tethered head depends on the rate of the power stroke, based on the landscape of stepping and the molecular details that the current experiment cannot easily access. Importantly, the present work will give further insights to resolve the recent experimental debate \cite{Coppin96PNAS,Nishiyama01NCB,Cross05Nature,Block07BJ} on the existence of a substep within the kinesin's 8-nm step. 
\\

\section*{Results}

\textbf{Topological constraints in stepping dynamics.}
Upon ATP binding to the empty MT-bound head (positioned at the binding site $o$ in Fig. 1A), 
the allosteric communication between the strained nucleotide binding pocket and the disordered neck-linker leads to the docking of the neck-linker to the 
$\beta7$ and $L10$ neck-linker binding motifs. As a result, 
the tethered head swings forward from one MT-binding site ($a$ in Fig. 1A) to the other ($e$ in Fig. 1A) 
while the MT-bound head remains at the same position (see SI Fig. 8 for the snapshots of a kinesin during the stepping).  
For the tethered head to bind to the MT-binding site, 
the ruggedly shaped kinesin head having specific MT-binding motifs should explore the MT surface and fit into the MT-binding site in a right orientation under topological constraints. 
The two neck-linkers (one from the MT-bound head, and the other from the tethered head) and the steric hindrance between the two heads restrain the search space available for the tethered head.  
The topology of MT also makes unique the association process between the tethered head and the MT-binding site. 
In the MT made of 13 PFs \cite{Downing02Structure} the lateral binding interfaces between the adjacent PFs are shifted by $\sim 0.95$ nm, which makes the MT helical (Fig. 1A). The distances between the diagonally located tubulins become asymmetric ($|\vec{od}|(=9.4$ $nm)<|\vec{of}|(=10.9$ $nm)$). 
The distance between the interfaced tubulins at the side ($|\vec{oc}|=|\vec{og}|=6.3$ nm) is shorter than the distance between the tubulins along the same PF ($|\vec{oe}|=8.0$ nm). 
In light of the contour length of a neck-linker (15 $aa\times 0.38$ nm$/aa$ $\approx 5.5$ nm), two fully extended neck linkers ($\approx 11$ nm) 
allow the tethered head to cover a wide range of the MT surface.
Since every MT-binding site on the MT equally interacts with the tethered head, one cannot totally rule out the possibility of binding to the neighboring PFs while the experiments suggest that the kinesin move straight along the MT \cite{Howard93JCB,Sheetz95BJ}. 
\\

\textbf{PMF between kinesin tethered head and MT.} 
During the disordered-to-ordered transition,
the length of disordered neck-linker decreases gradually. 
Depending on the length of the disordered neck-linker, there are substantial variations in the search space available for the tethered head (Fig. 2A and Fig. 3A).  
We sample the conformational space using the centroid of tethered head from the multiple trajectories with varying temperatures (see Methods for the energy function), and calculate the 2-D PMF between the tethered head and the MT, projected on the xy, xz-, yz-planes, at two extreme cases (see SI text for the computational details of PMF construction). 
One results from search processes under which the neck-linker of MT-bound head is practically in the zippered state ($\lambda=1$). We expect that this process is realized when the neck-linker zippering rate is faster than the rate of space exploration ($k_p\gg k_E$). Whereas, the other results from the case when the neck-linker is in disordered state throughout the search process ($k_p\ll k_E$, $\lambda=0$).  
Although the exact values of $k_p$ and $k_E$ are not known, we presume that the stepping dynamics should occur on a PMF linking between these two regimes. 

From the structures with $\lambda=1$, 
we find two major basins of attraction in the 3-D space (Fig. 2). One is the MT-binding site $e$ and the other is a broad basin ($S$) formed at the forward-right corner relative to the MT-bound head (Fig. 2C, $F_1(x,z)$). Between the two basins exists the free energy barrier of $\sim 1-2$ $k_BT$ in $F_1(x,z)$ or $F_1(x,y)$, and $\sim 4-5$ $k_BT$ in $F_1(y,z)$, depending on the projection. 
The basin $S$ stems from the large conformational degrees of freedom that 
the tethered head should explore before reaching the binding site $e$ (Fig. 2C, circles in blue).  
The energetic bias to the binding site $d$ (Fig. 2C, arrows in magenta) is also present, but is negligible ($< 1$ $k_BT$). 
With an increasing temperature the stability of basin $S$ relative to the 
binding site $e$ increases, which indicates that the nature of the broad basin is entropic (Fig. 2D). 
Note that the 1-D free energy profiles (Fig. 2B) projected from higher dimensional representation of free energy surface underestimates the barrier height by merging the various possible dynamic pathways. 

From the structures with 
$\lambda=0$, which is obtained by making repulsive the attractive neck-linker zipper contacts (the native contacts inside the green circles of Fig. 1B, see also Methods), we find that the most frequently visited binding site is the site $c$ that belongs to the adjacent PF (Fig. 3).
Explicit analysis on the bound complex shows that the binding of the tethered head to the site $c$ is not as complete as the one to the site $e$. 
Interestingly, the structure at $c$ has only $\sim 1/3$ of the interfacial native contacts (see Fig. 4). 
We find that a strained neck-linker induces a significant distortion in the $\alpha 6$ helix, preventing a full binding. \\

{\bf Modeling the stepping dynamics.}
The most straightforward strategy to monitor the real-time stepping dynamics of our model is to integrate the equation of motion for each coarse-grained unit of kinesin molecule in an overdamped environment. 
The inclusion of hydrodynamics to the simulations, which is essential to naturally retrieve a correct behavior of the translational diffusion for the tethered head (see SI text with SI Fig. 9), is however computationally too expensive for a model with more than 700 coarse-grained units. Thus, we take an alternative method to study the real-time dynamics by using the two PMFs obtained above. 

Conceptually, the stepping dynamics coupled to the power stroke is considered as a diffusional process with a moving reflecting boundary condition that restrains the search space, which is illustrated in Fig. 5A.
The blue shade and the red dots depict the energetic cost for extending the neck-linker and the multiple MT-binding sites, respectively. 
The progress of the neck-linker zippering is depicted by dotted lines with parameter $\lambda$.  
The reflecting boundary moves from $\lambda=0$ to $\lambda=1$ in a finite rate. Depending on the rate of the variation from $\lambda=0$ to $\lambda=1$ as well as the motility of the kinesin head, the dynamic pathway of binding process changes because of the influence of the sideway binding site $c$. 
The rate of variation from $\lambda=0$ to $\lambda=1$ is defined as 
$k_p=\tau_p^{-1}=d\lambda/dt$, i.e., $\lambda=k_pt$ with $0\leq t\leq \tau_p$.

We propose that the PMF time-dependently linking between the two extreme cases be modeled using the following ansatz adapting the dual-Go potential \cite{Best05Structure,Hyeon06PNAS},
\begin{equation}
\beta F(x,y,z,t)=\left\{ \begin{array}{ll}
-\log{\left[\left(1-\frac{t}{\tau_{p}}\right)e^{-\beta F_0(x,y,z)}+\frac{t}{\tau_{p}}e^{-\beta F_1(x,y,z)}\right]}&\mbox{$(0\leq t\leq \tau_p)$}\\
\beta F_1(x,y,z)&\mbox{$(t\geq \tau_p)$}\end{array}\right.
\label{eqn:tPMF}
\end{equation}
where the switching rate between $F_0(x,y,z)$ and $F_1(x,y,z)$ is controlled by $\tau_p(=k_p^{-1})$. 
We perform Brownian dynamics simulations of a quasi-particle, representing the 
centroid of tethered head, on the ``time-dependent'' PMF, $F(x,y,z,t)$, from the initial position of the head at $(x_i,y_i,z_i)=(-6.0,4.0,-2.0)$ $nm$. We update the position at time $t$ using 
$\vec{r}(t+\Delta t)=\vec{r}(t)-D^{eff}_K\vec{\nabla}F(x,y,z,t)\Delta t/k_BT+\vec{R}(t)$
where $\vec{R}(t)$ is a vector of Gaussian random number satisfying $\langle \vec{R}\rangle=0$ and $\langle \vec{R}\cdot\vec{R}\rangle=6D_K^{eff}\Delta t$.
We choose $D^{eff}_K=2$ $\mu m^2/s$ for the effective diffusion coefficient of the tethered head to reproduce a similar time scale as the one in the recent SM experiment by Yanagida and coworkers \cite{Nishiyama01NCB}, where they monitored the stepping dynamics of kinesin in the time resolution of $\sim 20$ $\mu s$ and suggested that the kinesin take substeps at 
the displacement of $\Delta x\sim 4$ $nm$.

Simulating the dynamic trajectories by varying the $\tau_p$, we count the number of trajectories trapped in the binding site $c$ (Fig. 5B). The fraction of trajectories directly reaching the site $e$ decreases almost exponentially as $\tau_p$ increases. We find that when $\tau_p\lesssim 20$ $\mu s$, 90 \% of the stepping dynamics occur without being trapped to the intermediate. 
The signature of such an intermediate, manifested as a ``substep'' in the averaged trace, is captured only when $\tau_p>20$ $\mu s$ (Fig. 5C). 
Note the similarity between the patterns of averaged time traces by Yanagida and coworkers (see Figure 4a in Ref.\cite{Nishiyama01NCB}) and the present results plotted in Fig. 5C. 
Interestingly, the averaged trace for $\tau_p=20$ $\mu s$ fits to the double exponential function (Fig. 5C) where the time scale of the fast phase corresponds to that of the power stroke, and the slow phase represents the rate process to overcome the barrier from broad basin to the target basin shown in $F_1(x,y,z)$.  
\\

\textbf{Partial unfolding (cracking) of structure facilitates the binding process.}
In the above sections, we studied the association between the tethered kinesin head and the MT-binding site, using the centroid of the tethered head.  
When we probe the kinesin conformation and the binding interface using native contacts,
we observe partial unfolding \cite{Miyashita03PNAS} of the MT-binding motifs in the kinesin head at the transition state ensemble prior to the complete binding. The flexibility of the structure eases the binding process by reducing the entropic barrier to overcome, which is well known phenomena in protein-protein or protein-DNA association process \cite{Shoemaker00PNAS,Dyson05NRMCB,Levy07JACS}. 
To quantify the degree of cracking in the kinesin structure during binding, we use the fraction of native contacts for the MT-binding motifs of kinesin ($Q_p$) and the fraction of interfacial native contacts between the kinesin and the MT ($Q_{int}$) (Fig. 6). In an exemplary trajectory (Fig. 6A) starting from $Q_p\approx 0.75$, 
the MT-binding motifs are disrupted to $Q_p\approx 0.65$. When the kinesin is completely bound to the MT, $Q_p\approx 0.82(\pm 0.08)$. 
By collecting the configurations and applying the histogram re-weighting technique (see SI text), we obtain the 2-D free energy surface $F(Q_p,Q_{int})$. 
Figs. 6C, 6D show that the $Q_p$ value of the kinesin-MT complex is greater than the value of the transition state ensemble. At higher temperature, 
the trend of cracking prior to the binding is more pronounced, showing a downward curvature in the pathways connecting the separated molecules and the complex. 
Interestingly, the 1-D free energy profile $F(Q_{int})$ (Fig. 6B) shows that the free energy barrier for the binding is $\sim 6$ $k_BT$, which is higher than the ones measured as function of spatial reaction coordinates in the Fig. 2.
\\

\section*{Discussions}
To study the dynamics of kinesin's stepping motion, we first made full use of the topological information available from the structures of kinesin and MT to build 
the PMF, and second considered the whole stepping dynamics as a ``rectified diffusional motion'' \cite{Peskin93BJ,Yanagida05NatureChemBiol} by envisioning the 
power stroke as a moving reflecting boundary condition for the diffusional motion of the tethered head. 
Although the level of our description is solely hinged on the native topology, lacking in the chemical details, 
several key issues on the stepping dynamics can be discussed in a semiquantitative fashion.

(i) The PMF between the tethered head and the MT showed that the leftward diagonal stepping ($\vec{of}$) is forbidden because of structural constraints (see Figs. 2 and 3).  
Also, the alternation between the sideway stepping ($\vec{oc}$ stepping) and the parallel stepping would generate a helical path in the long run, which contradicts to the previous experimental findings \cite{Sheetz95BJ}. 
Thus, the likelihood for using two parallel PFs is ruled out. 
An interesting finding in this study is that the intermediate structure trapped into the sideway binding site $c$ has only $\sim 30$ \% stability compared with 
the correctly bound complex at $e$. 
We expect that under the action of power stroke, the intermediate state found for $k_p\ll k_E$ becomes further destabilized and readily loses its binding with the site $c$. 
A direct comparison between $F_0(x,y,z)$ and $F_1(x,y,z)$ at $(x,y,z)\approx (0,4,5)$ nm indicates that the binding site $c$ is destabilized by $\gtrsim 5$ $k_BT$ upon $\lambda=0\rightarrow \lambda=1$. 

(ii) 
With the speed limit of protein folding rate $\sim \mathcal{O}(1)$ $(\mu s)^{-1}$ \cite{ThirumJPI,OnuchicCOSB04,Eaton04COSB} and the number of amino acids consisting of neck-linker $N\sim(12-15)$, one can roughly estimate the rate of power stroke using the scaling relation for the folding rate of proteins with $N$ ($k_F\sim k_F^0\exp{(-1.1 N^{1/2})}$ with $(k_F^0)^{-1}\sim 0.4$ $\mu s$ \cite{ThirumJPI,LI04Polymer}, or $k_F\sim k_F^0\exp{(-0.36N^{2/3})}$ with $(k_F^0)^{-1}\sim 8$ $\mu s$ \cite{Bryngelson90Biopolymers}). The agreement of $k_F^{-1}(\sim 20-70$ $\mu s)$ with $k_p^{-1}(\sim 20$ $\mu s)$ shows that the power stroke is closely connected with the conformational change of neck-linker whose activation barrier is estimated as 
$\Delta G^{\ddagger}/k_BT(\approx 1.1 N^{1/2}$ or $0.36 N^{2/3})\sim 2-4$. 

(iii) Given the diffusion constant ($D_K^{eff}$) and the approximate shape of PMF, one can estimate $k_E$, the exploration rate over the PMF, by approximating the kinesin's motion as a Brownian motion in a harmonic potential ($F(x)\sim 1/2\times kx^2$, i.e., Ornstein-Uhlenbeck process), whose conditional probability is solved as follows \cite{Uhlenbeck30PR}. 
\begin{equation}
W(x,t|x_0)=\left[\frac{2\pi k_BT}{k}(1-e^{-2t/\tau_E})\right]^{-1/2}\exp{\left(-\frac{(x-x_0e^{-t/\tau_E})^{2}}{\frac{2k_BT}{k}(1-e^{-2t/\tau_E})}\right)},
\end{equation}
where $\tau_E=k_BT/D_K^{eff}k$. When $t\gg \tau_E$, $W(x,t|x_0)\rightarrow P_{eq}(x)$, $\langle x\rangle_{eq}\rightarrow 0$, and 
$\langle (\delta x)^2\rangle_{eq}\rightarrow k_BT/k$. 
Since $k\approx 0.02$ $k_BT/nm^2$ from the fit of 1-D $F_0(x)$ to a quadratic potential (Fig. 3C), and $D_K^{eff}=2$ $\mu m^2/s$, one gets $\tau_E=25$ $\mu s$. 
The lower bound for the rate of power stroke ($k_p$) is similar to the upper bound for the exploration rate of harmonic potential ($k_E=\tau_E^{-1}$).  
If the PMF switching is too fast, the molecules sample only the subregion of the landscape, reflecting the signature of the far-from-equilibrium dynamics. 

(iv) The experimentally measured time traces for the rising phase, averaged over the different stepping time scales, by Yanagida and coworkers \cite{Nishiyama01NCB} showed the signatures of intermediates.
The comparison between the data by Yanagida and coworkers' and our time traces generated over varying $k_p$ suggests that the $k_p$ distribute broadly, given by a distribution of rate constant $g(k_p)$. 
Because of the molecular origin of the power stroke, associated with a complex energy landscape representing the molecular architecture, 
it is natural to speculate that the stepping dynamics occurs via the kinetic partitioning mechanism \cite{HyeonBC05}.
The multiple basins (the binding sites $c$, $d$, and entropic basin $S$) in the PMF suggests that kinesins reach the target binding site $e$ along multiple parallel pathways. 
Not dividing the time traces into the different classes like the way Yanagida and coworkers adopted \cite{Nishiyama01NCB}, one should be able to fit the full ensemble average of the time traces to the finite number of multi-exponential function as $\langle x\rangle (t)=\int_0^{\infty} dk_pg(k_p)e^{-k_pt}$ \cite{Zwanzig90ACR}. 
Although a visual signature of intermediate (a substep in the averaged time trace) is masked if the contribution from the fast kinetics ($\tau_p<20$ $\mu s$) is dominant, 
a careful statistical analysis of $\langle x\rangle (t)$ averaged over the entire time traces would give a glimpse to the $g(k_p)$ that encodes the underlying energy landscape associated with the stepping dynamics. 
\\

\section*{Methods}
To generate the configurations of kinesin on the MT, we performed hybrid Monte-Carlo/Molecular Dynamics simulations by adapting an ADP-complexed crystal structure of rat kinesin dimer (PDB code : 3kin) on the 13-PF MT structure (see SI text for further details). 
The 13-PF structure is built with multiple tubulin dimers fitted to the Downing and coworkers' 8-\AA\ resolution electron density map \cite{Downing02Structure}.
\\

\textbf{Energy function.}
The energy function for the kinesin/MT system is modeled as  
$V_{\mathrm{tot}}=V_{\mathrm{K}}+V_{\mathrm{K\cdot MT}}$ where $V_{\mathrm{K}}(=V_{\mathrm{K}}^{\mathrm{B}}+V_{\mathrm{K}}^{\mathrm{T}}+V_{\mathrm{K}}^{\mathrm{BT}})$ is the energy function for the kinesin dimer and $V_{\mathrm{K\cdot MT}}$ is the interaction between the kinesin and the MT. 
B and T denote the MT-bound head and the tethered head, respectively. $V_{\mathrm{K}}^{\mathrm{BT}}$ describes the interaction between the B and T head. 

To design each kinesin monomer we use the self-organized polymer model \cite{Hyeon06Structure,Hyeon06PNAS},  
\begin{align}
V_{\mathrm{K}}^{\mathrm{\xi}}&=\sum_{i=1}^{N_{\xi}-1}{\left(-\frac{k}{2}R_o^2\log{\left[1-\frac{(r_{i,i+1}-r^o_{i,i+1})^2}{R_o^2}\right]}\right)}\nonumber\\
&+\sum_{i=1}^{N_{\xi}-3}\sum_{j=i+3}^{N_{\xi}}\varepsilon_h(i,j)\left[\left(\frac{r^o_{ij}}{r_{ij}}\right)^{12}-2\left(\frac{r^o_{ij}}{r_{ij}}\right)^6\right]\Delta_{ij}\nonumber\\ 
&+\sum_{i=1}^{N_{\xi}-2}\varepsilon_l\left(\frac{\sigma}{r_{i,i+2}}\right)^6+\sum_{i=1}^{N_{\xi}-3}\sum_{j=i+3}^{N_{\xi}}\varepsilon_l\left(\frac{\sigma}{r_{ij}}\right)^6(1-\Delta_{ij})
\label{eqn:SOP}
\end{align}
where $\xi=$B or T. 
The first term models the chain connectivity with $k=100$ $kcal/(mol\cdot$\AA$^2)$, $R_o=2$ \AA. The second term decribes the attraction between the native contact pairs $i$ and $j$. The native pairs are defined using $\Delta_{ij}=1$ for the residue pairs within $R_c=8$ \AA\ at the native structure, and $r_{ij}^o$ is the corresponding distance. 
$\varepsilon_h(i,j)$ controls the strengths of native pairs.  
We choose $\varepsilon_h(i,j)=1.8$ kcal/mol for the residue $i$ and $j(=i+3)$ within an $\alpha$-helix, and  $\varepsilon_h(i,j)=1.2$ $kcal/mol$ for all other cases, e.g., $\beta$-sheets, loops, and residue pairs between the secondary structural elements.
The third term prevents the volume overlap between the residues $i$ and $i+2$. 
The last term is for the repulsive potential between the non-native pairs, modeled using $\varepsilon_l=1$ $kcal/mol$ and $\sigma=3.8$ \AA.  
We design the L12 loop, disordered in the crystal structure, as a self-avoiding chain, employing the same Hamiltonian as Eq.\ref{eqn:SOP} with $\Delta_{ij}=0$ for $i=240-255$ and all other $j$, or vice versa, which makes the L12 loop neutral to other parts of kinesin monomer (see also the SI text). 
The coiled-coil interaction responsible for dimerization between two neck-helices (residues 341$-$370) is modeled using 
\begin{align}
V_{\mathrm{K}}^{\mathrm{BT}}&=\sum_{i=1}^{N_{\mathrm{B}}}\sum_{j=1}^{N_{\mathrm{T}}}\varepsilon_h^{\alpha\alpha}\left[\left(\frac{r^o_{ij}}{r_{ij}}\right)^{12}-2\left(\frac{r^o_{ij}}{r_{ij}}\right)^6\right]\Delta_{ij}^{\alpha\alpha}+\sum_{i=1}^{N_{\mathrm{B}}}\sum_{j=1}^{N_{\mathrm{T}}}\varepsilon_l\left(\frac{\sigma}{r_{ij}}\right)^6(1-\Delta_{ij}^{\alpha\alpha}),
\label{eqn:BT}
\end{align}
with $\varepsilon_h^{\alpha\alpha}=1.2$ $kcal/mol$. 
We defined $\Delta_{ij}^{\alpha\alpha}=1$ only for ``$i,j\geq 341$'' with $R^o_{ij}<R_c$, otherwise $\Delta_{ij}^{\alpha\alpha}=0$, 
so that the two motor domains other than neck-helix repel each other. 
Because the two monomers are identical as a fold, we should impose the ``same'' topological bias to the both heads, which is achieved by setting $\Delta_{ij}(T)=\Delta_{ij}(B)$ and $r_{ij}^o(T)=r_{ij}^o(B)$ for all $i$ and $j$. This condition was used when we have previously studied the role of internal strain on the regulation mechanism of kinesin dimer \cite{Hyeon07PNAS}. 
To study the stepping dynamics in particular, 
however, we want the neck-linker of the tethered head in a disordered state, 
so that the tethered head can search and reach the next MT-binding site. 
This is realized by making the neck-linker (residues 327-338) of the tethered head always repulsive to the residues constituting the neck-linker binding site (see Fig. 1B). 
To generate the kinesin configurations with $\lambda=1$ and $\lambda=0$, 
we retain ($\lambda=1$) or discard ($\lambda=0$) the neck-linker zipper contacts (contacts inside green circle in Fig. 1B) of the MT-bound head. 

The interaction between the kinesin and the MT is designed using 
\begin{align}
V_{\mathrm{K\cdot MT}}&=\sum_{i=1}^{N_{\mathrm{K}}}\sum_{k=1}^{N_{\mathrm{MT}}}\left[\varepsilon_h^{K\cdot MT}\left\{\left(\frac{r^o_{ik}}{r_{ik}}\right)^{12}-\chi_{ik}\left(\frac{r^o_{ik}}{r_{ik}}\right)^6\right\}+l_Bk_BT\frac{z_iz_k}{r_{ik}}e^{-r_{ik}/l_D}\right]\Delta^*_{ik}\nonumber\\
&+\sum_{i=1}^{N_{\mathrm{K}}}\sum_{k=1}^{N_{\mathrm{MT}}}\left\{\varepsilon_l^{K\cdot MT}\left(\frac{\sigma}{r_{ik}}\right)^6+l_Bk_BT\frac{z_iz_k}{r_{ik}}e^{-r_{ik}/l_D}\right\}(1-\Delta^*_{ik}).
\label{eqn:KMT}
\end{align}
$N_{\mathrm{MT}}$ is the number of all the residues, belonging to the tubulin heterodimers, that are within the interaction range from kinesin when the MT-bound head is poised at the central tubulin (o) (see Fig. 1A). 
The topological information of the binding interface between the MT-bound head and the central tubulin (o) is replicated to other surrounding tubulins (a-h) using $\Delta_{ik}^*$, so that the tethered head feels the identical potential on every MT-binding site.  
The native contacts between the kinesin and MT are defined between the $i$, $k$ pairs within $R_c^{\mathrm{K\cdot MT}}=9$ \AA. 
$\Delta_{ik}^*=1$ for the native pairs, and $\Delta_{ik}^*=0$ otherwise. 
In addition to the non-bonded interaction ($\varepsilon_h^{K\cdot MT}=\varepsilon_l^{K\cdot MT}=1$ $kcal/mol$), the electrostatic interaction between the kinesin and the MT is considered because of the large amount of net negative charge ($-35$ $e$) in each tubulin unit. 
To preserve the native contact distance as the one in the crystal structure even with electrostatic potentials, we adjust the parameter $\chi_{ik}$ in the Eq.\ref{eqn:KMT} by choosing $\chi_{ik}=2+(l_Bk_BT/6\varepsilon_h^{K\cdot MT})[1/r_{ik}^o+1/l_D]e^{-r_{ik}^o/l_D}$. 
At $r=r_{ik}^o$, the energy for the native pair with the opposite charges is given as 
$V(r_{ik}^o)=-\varepsilon_h^{K\cdot MT}-l_Bk_BT\left(7/6r_{ik}^o+1/l_D\right)e^{-r_{ik}^o/l_D}$.
The strength of electrostatics is controlled by the salt concentration $c$, that determines the Debye screening length $l_D=(8\pi l_B c)^{-1/2}\approx (3/\sqrt{c})$ \AA\ where $l_B=7$ \AA\ and $c$ is in the unit of M ($mol/l$). 
For a pair formed at $r_{ik}^o\approx 8$ \AA, the energy stabilization due to the electrostatics is $l_Bk_BT\left(7/6r_{ik}^o+1/l_D\right)e^{-r_{ik}^o/l_D}\approx 0.14$ $kcal/mol$ at $c\approx 1$ M, but the same value becomes $\approx 1.4$ kcal/mol at $c\approx 0.1$ M, which is comparable to $\varepsilon_h=1.0$ kcal/mol. 
At physiological condition, $c\sim (0.1-0.2)$ M for monovalent salt. A careful consideration is, however, required in choosing the $c$ near the MT surface because of the counterion condensation \cite{Manning69JCP}. Near the highly charged rod, 
the counterion concentration can be much higher than that of bulk. 
Therefore, we choose $c=1$ M, which makes the electrostatics negligible, throughout the whole simulations (see SI text with SI Fig. 10 for the details). 
Eq.\ref{eqn:KMT} accomodates nonspecific interactions due to electrostatics even for the non-native contacts.
\\ 

\textbf{Sampling the free energy surface. }
We combine Monte Carlo (MC) and molecular dynamics (MD) simulations to efficiently sample the kinesin configurations on the MT. 
We first generate a ensemble of initial configurations by both pivoting \cite{BishopJCP91} a random position of neck-linkers (residues 327-339 for the MT-bound head, residues 328-337 for the tethered head) and translating the center of position of the tethered head. 
The acceptance of 
a trial move is decided by standard Metropolis criteria with the potentials defined in Eqs. \ref{eqn:SOP}, \ref{eqn:BT}, and \ref{eqn:KMT}. 
The subsequent hybrid MC-MD simulations are performed from a host of initial kinesin configurations. During the 1000 simulation steps, we perform the MC simulations for the first 50 steps, and integrate the subsequent 950 steps using a velocity-Verlet algorithm. The kinesin configuration is collected every 1000 step. 
\\

\section*{Acknowledgements}
We thank Dr. Stefan Klumpp for insightful conversation, and Dr. Ken Downing for kindly providing the coordinate of 13-PF MT.  
This work was funded by NSF grant MCB-054396 and by the NSF-sponsored Center for Theoretical Biological Physics (Grants PHY-02165706 and PHY-02255630). 

\clearpage
\section*{Supporting Information}
\textbf{Preparation of kinesin and MT structure : }
The simulations of kinesin were performed, referenced to an ADP-complexed crystal structure of rat kinesin dimer (PDB code : 3kin) in which 
the neck-linkers of both monomers are in an ordered state.
For the completeness, using a self-avoiding chain, we filled the gap of the missing residues 240-255, whose sequence is s\textbf{k}tgaegavld, corresponding to the L12 loop in the crystal structure. 
Unlike the monomeric kinesin, KIF1A, whose L12 loop contains many lysine residues implying an important role in the motility by interacting with negatively charged E-hook of the tubulin \cite{Hirokawa00PNAS}, the L12 loop of rat kinesin has only one lysine. Presumably, the electrostatic between the L12 loop and MT surface is not as important as the one in KIF1A. 
As an initial configuration, 
one of the kinesin monomers was placed on a tubulin binding site of the MT, 
and the other monomer is tethered to the MT-bound monomer via coiled-coil association but away from the direct MT interaction range.  The topological information of binding interface between kinesin and tubulin was acquired using the Hirokawa and coworkers' KIF1A and tubulin complex \cite{HirokawaSCI04} (see also Ref. \cite{Hyeon07PNAS} for the detailed procedure).\\ 

{\bf Computation of two-dimensional potential of mean force (PMF) : }
The multiple histogram reweighting technique \cite{SwendsenPRL88,KumarJCC1992} was adopted 
to compute the two-dimensional PMF between the tethered kinesin head and MT at temperature $T$. For example, the 2-D PMF can be obtained at arbitrary values of $T$ if the conformational states are well sampled over a range of $T$ values.
The probability of finding the kinesin head at position $(x,z)$ at temperature $T$ is given by
\begin{equation}
P(x,z)(T)=\frac{\sum_{E}e^{-E/T}\frac{\sum_{k=1}^Kh_k(E,x,z)}{\sum_{k=1}^K n_ke^{(F_k-E)/T_k}}}{\sum_{E,x,z}e^{-E/T}\frac{\sum_{k=1}^Kh_k(E,x,z)}{\sum_{k=1}^K n_ke^{(F_k-E)/T_k}}}
\label{eq:WHAM}
\end{equation}
where $K$ is the number of histograms,
$h_k(E,x,z)$ is the number of states between ($E$,$E+\delta E$), ($x$,$x+\delta x$), and ($z$,$z+\delta z$) in
the $k$-th histogram, $n_k=\sum_{E,x,z}h_k(E,x,z)$,
$T_k$ is the temperature in the simulations used to generate the
$k-$th histogram. 
The free energy, $F_k$, that is calculated self-consistently, satisfies
\begin{equation}
e^{-F_r/T_r}=\sum_{E,x,z}e^{-E/T_r}\frac{\sum_{k=1}^Kh_k(E,x,z)}{\sum_{k=1}^Kn_ke^{(F_k-E)/T_k}}.
\label{eq:WHAM2}
\end{equation}
The self-consistent equation for $Z_k\equiv e^{-F_k/T_k}$ converge to the final values of $\{Z_k\}$ starting from $Z_k=1$ ($k=1,2,\ldots,K$). 
Using the hybrid MC/MD simulations,
we sampled the conformational states over the range of $295$ $K$$<T<356$ $K$.
Once $P(x,z)(T)$ is obtained, the 2-D PMF as a function of $(x,z)$ is given by 
\begin{equation}
\Delta F(x,y)(T)=-k_BT\log{P(x,z)(T)}. 
\label{eq:freeprofile}
\end{equation}
$\Delta F(x,y)$, $\Delta F(y,z)$, and $\Delta F(Q_{int},Q_p)$ are similarly obtained, and the 1-D PMFs are easily reduced from the 2-D PMFs. \\

\textbf{Estimate of the translational diffusion constant of kinesin motor domain : }
When performing the Brownian dynamics simulation, we decided the diffusion constant of monomer using Stokes-Einstein relation, 
\begin{equation}
D_o=\frac{k_BT}{6\pi\eta a}, 
\end{equation}
where $\eta$ is the viscosity of water ($\approx 1cP=10^{-3}N/m^2\cdot sec$), and $a$ is the hydrodynamic radius of residue ($a\approx 0.19+0.14$ nm), so that the monomer diffusion constant at $T=300$ K is $D_m\approx 578$ $\mu m^2/sec=5.8\times 10^{-6}cm^2/sec$. Similar calculation for the spherical object of radius $R\approx 4$ nm (approximately the size of kinesin head) leads to $D_K\approx 5.5\times 10^{-7}cm^2/sec=55$ $\mu m^2/n sec$. \\

\textbf{Brownian dynamics with hydrodynamics :}
To simulate the real time kinetics of kinesin's swiveling motion using a Brownian dynamics of coarse-grained model, the inclusion of hydrodynamics is extremely 
important to obtain a correct time scale for the translational diffusion of whole object. 
Without hydrodynamics, the translational diffusion constant of a protein ($D_K$) 
that is coarse-grained by the N beads is scaled as $D_o/N$, where $D_o$ is the diffusion constant of the single bead.  
To naturally satisfy the Stokes-Einstein relation for the translational diffusion constant,
the Brownian dynamics simulations requires the inclusion of hydrodynamic interaction. 
When off-diagonal elements of diffusional tensor is included and preaveraged, the diffusion constant of whole object $D_K$ scales as \cite{KremerJCP93,Rotne69JCP} 
\begin{equation}
D_K=\frac{D_o}{N}+\frac{k_BT}{6\pi\eta R_H}\sim \frac{D_o}{N}+\frac{D_o}{N^{\nu}}\rightarrow \frac{D_o}{N^{\nu}}. 
\end{equation}
$D_K\sim D_o/N^{\nu}$ is the correct scaling relative to the $D_o$. 
Without hydrodynamics, the translational diffusion constant is significantly underestimated especially 
when $N$ is large. 
However, it is also challenging to simulate the Brownian dynamics with hydrodynamic interaction. 
The Langevin equation under multidimensional space is 
\begin{equation}
{\bf r}_i(t+\Delta t)-{\bf r}_i(t)=\frac{1}{k_BT}\sum_{j}{\bf D}_{ij}\cdot {\bf F}_j+\sqrt{2}\sum_j{\bf B}_{ij}\cdot {\bf n}_j(t)
\end{equation} 
with $\left<{\bf n}_i(t)\right>=0$, $\left<{\bf n}_i(t)\cdot {\bf n}_i(t)\right>=\delta_{ij}\delta(t-t')$ and ${\bf D}={\bf B}{\bf B}^T$. The diffusion tensor ${\bf D}$ should be positive definite i.e. $\sum_{ij}{\bf F}_i\cdot {\bf D}_{ij}\cdot {\bf F}_j>0$ for all ${\bf F}\neq 0$, then ${\bf D}$ is deomposed into 
${\bf B}$ and ${\bf B}^T$ using the Cholesky decomposition. 
For the hydrodynamic diffusion tensor, Rotne-Prager diffusion tensor \cite{Rotne69JCP} is used. 
By averaging the square displacement of many different trajectories, we 
show in Fig.\ref{hydro} how the explicit inclusion of hydrodynamics can alter the translational diffusion constant of the kinesin head domain. 
The value of $D(hydro)=19.9$ $\mu m^2/sec$ is closer to the expected value for the $D_K(\approx 55$ $\mu m^2/sec$. See above) that can be roughly estimated from the Stokes-Einstein relation. 
\\ 

\textbf{Consideration of electrostatics $-$ Manning counterion condensation around the microtubule : }
The microtubule is viewed as a cylindrical object with many charges. 
Unlike other thin polyelectrolyte such as RNA, ss-DNA and ds-DNA that can adapt its conformation depending on the salt concentration, the MT can serve as an excellent example to apply the Manning condensation theory for the cylindrical object. 
Because each tubulin hetero-dimer contains $\sim 35e$ negative charge, the 
line charge density of microtubule is computed as 
\begin{equation}
d\approx \frac{(13\times 35 e)}{80\mathrm{\AA}}\approx 5.6e/\mathrm{\AA}. 
\end{equation}
The large Manning condensation parameter \cite{Manning69JCP} $\xi=l_B/b(\approx 7.1\AA/(1/5.6)\AA)\approx 39.8$ suggests that a drastic counterion condensation should occur around the microtubule to make $\xi\approx 1$. This requires $n\sim 34.1e$ monovalent positive counterion, which is estimated from $b^*\approx l_B$, i.e., $\frac{82}{13\times(35-n)}\approx 7.1$, should condense to the surface of tubulin. 
The numerical solution of nonlinear Poisson-Boltzmann equation with $\phi'(a)=-\frac{2\xi}{a}$ and $\phi(\infty)=0$
\begin{equation}
\nabla^2\phi=\kappa^2\sinh{\phi},
\end{equation}
where $\phi(r)\equiv e\psi(r)/k_BT$ and $\kappa=\sqrt{8\pi l_B c}$,  
determines the ion distribution around the charged cylinder. 
With $\phi(r)$ value, the number of positive and negative ions per $l_B$ around the cylinder can be computed using $n_+(r)=n_{\infty}e^{-\phi}$ and $n_-(r)=n_{\infty}e^{\phi}$, respectively. 
The net condensate charge number $q(r)=(n_+(r)-n_-(r))$ provides the thickness of 
condensate ion by imposing the Manning condensation condition, $q(R_M)\times l_B=1$. For the parameter $a=12.5$ $nm$, $c=150$ $mM$, the counterion condensation occurs at $R_M\approx 16.7$ $nm$ (see Fig.\ref{condensation}). 
About 4 $nm$ condensate counterion layer is 
formed around the cylinder. Unless the interfacial binding takes place and expulges the counterion condensated near the binding site, the electrostatic 
potential due to the microtubule charge does not significantly affect the 
swiveling dynamics of kinesin head. 
The widely-accepted notion of ``electrostatic steering'' in the context of protein-protein association dynamics is not clear for the object on highly charged but effectively neutralized cylindrical surface. 
It has been shown that the processivity of the kinesin is affected by the salt concentration \cite{Gilbert95Nature}
\\

\clearpage
\begin{figure}[ht]
\includegraphics[width=7.00in]{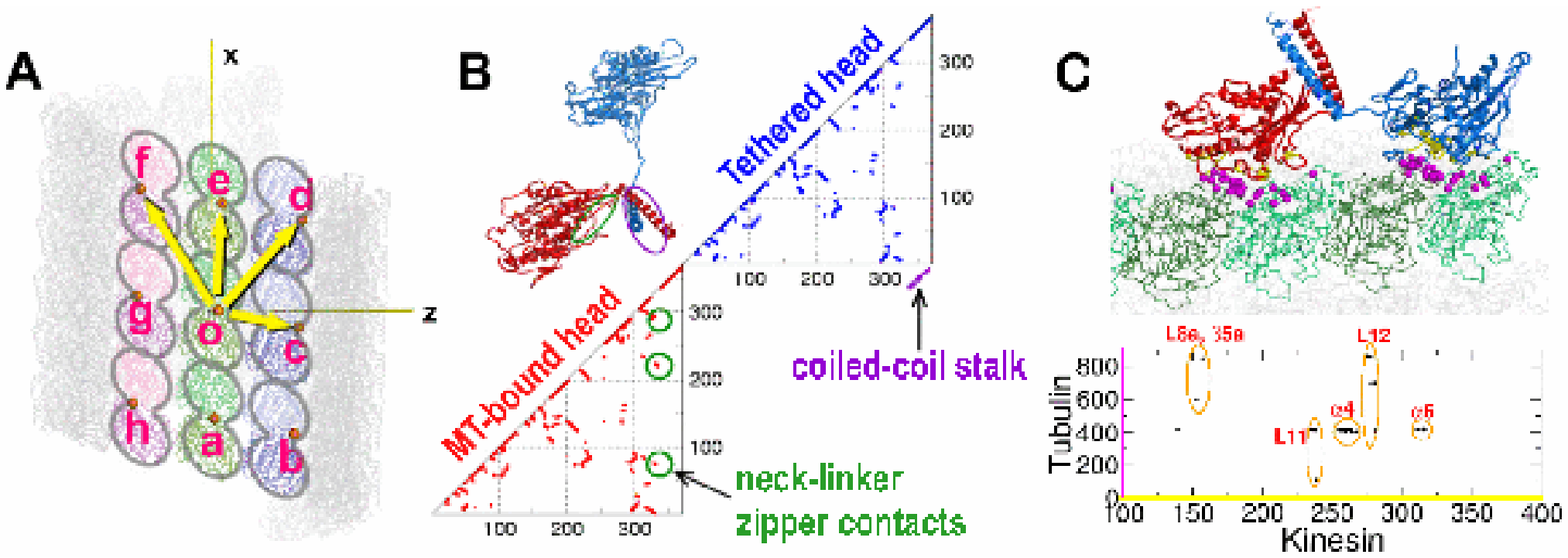}
\caption{Topology of MT, kinesin, and their interface.
{\bf A}. Geometry of MT surface showing the distances between the neighboring binding sites.
$|\vec{oa}|=|\vec{oe}|=8.0$ nm, $|\vec{ob}|=|\vec{of}|=10.9$ nm, $|\vec{oc}|=|\vec{og}|=6.3$ nm, $|\vec{od}|=|\vec{oh}|=9.4$ nm.
The distances are measured using the position of the residue 400 at each $\alpha$-tubulin subunit.
{\bf B}. The native contact map of kinesin model. The native bias between the neck-linker and its neck-linker binding site (neck-linker zipper contacts inside the green circles) is retained for the MT-bound head to have the neck-linker ordered ($\lambda=1$) while it is removed for the tethered head to make the neck-linker of the tethered head disordered.
For structures with a disordered neck-linker in the MT-bound head ($\lambda=0$), the neck-linker zipper contacts are made repulsive. The contacts for the coiled-coil are colored purple.
{\bf C}. Native contacts between the residues of kinesin (yellow) and MT-binding sites (magenta).
\label{Topology}}
\end{figure}
\begin{figure}[ht]
\includegraphics[width=6.80in]{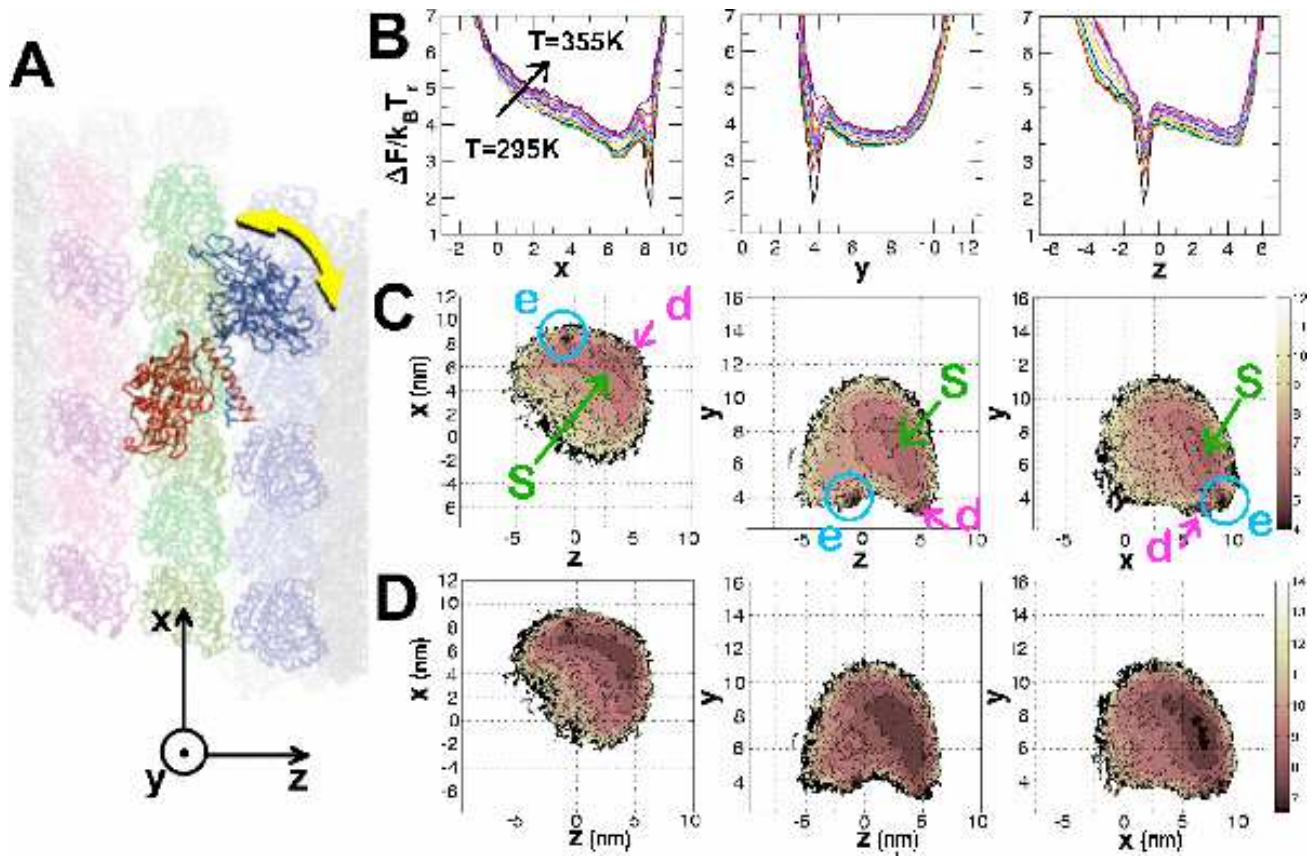}
\caption{PMF between the tethered head of kinesin and the MT surface for $\lambda=1$.
{\bf B}. 1-D PMFs in x-, y-, z- projections are shown at various temperatures. {\bf C}. 2-D PMFs in xz-, yz-, xy- projections at $T=295$ K.
The binding sites $e$, $d$, and entropic basin $S$ are marked with circles and arrows.
{\bf D}. 2-D PMFs at $T=355$ K.
\label{Free_zipped}}
\end{figure}
\begin{figure}[ht]
\includegraphics[width=6.80in]{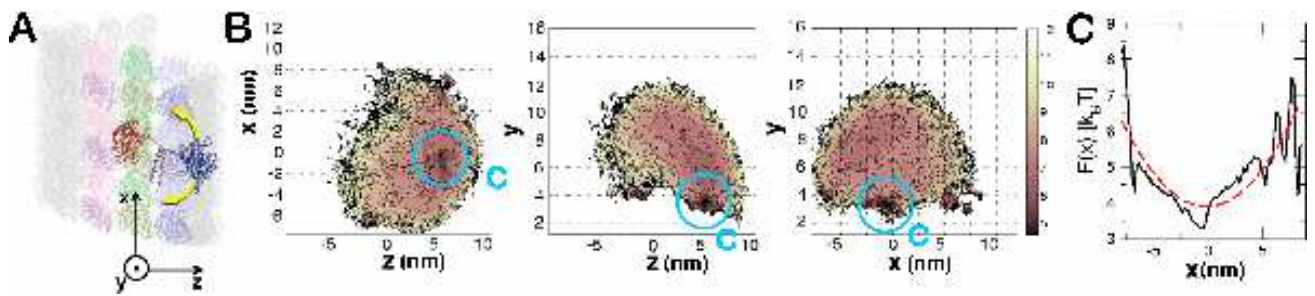}
\caption{PMF between the tethered head of kinesin and the MT surface for $\lambda=0$.
{\bf B}. 2-D PMFs in xz-, yz-, xy- projections at $T=295$ K.
The binding site $c$, which results in an intermediate state, is marked with circles.
{\bf C}. 1-D free energy profile $F_0(x)$ is fitted by a harmonic potential $1/2$ $kx^2$ where $k\approx 0.02$ $k_BT/nm^2$ (dashed line).
\label{Free_nozipped}}
\end{figure}
\begin{figure}[ht]
\includegraphics[width=6.80in]{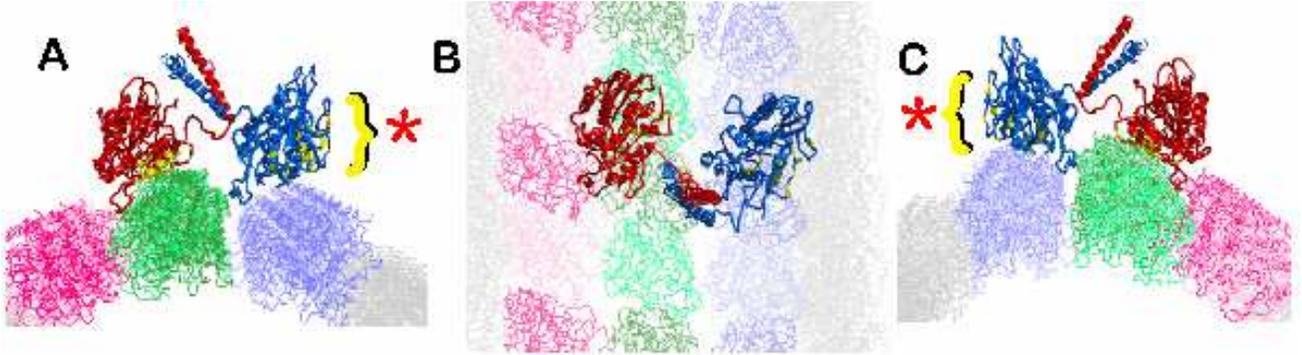}
\caption{The conformation of kinesin dimer (view from the back ({\bf A}), top ({\bf B}), and front ({\bf C})) when the tethered head is trapped at binding site $c$. The tethered head are partially bound to the site $c$, using only $1/3$ of the native contacts.
\label{sidewaybindingconf}}
\end{figure}
\begin{figure}[ht]
\includegraphics[width=5.00in]{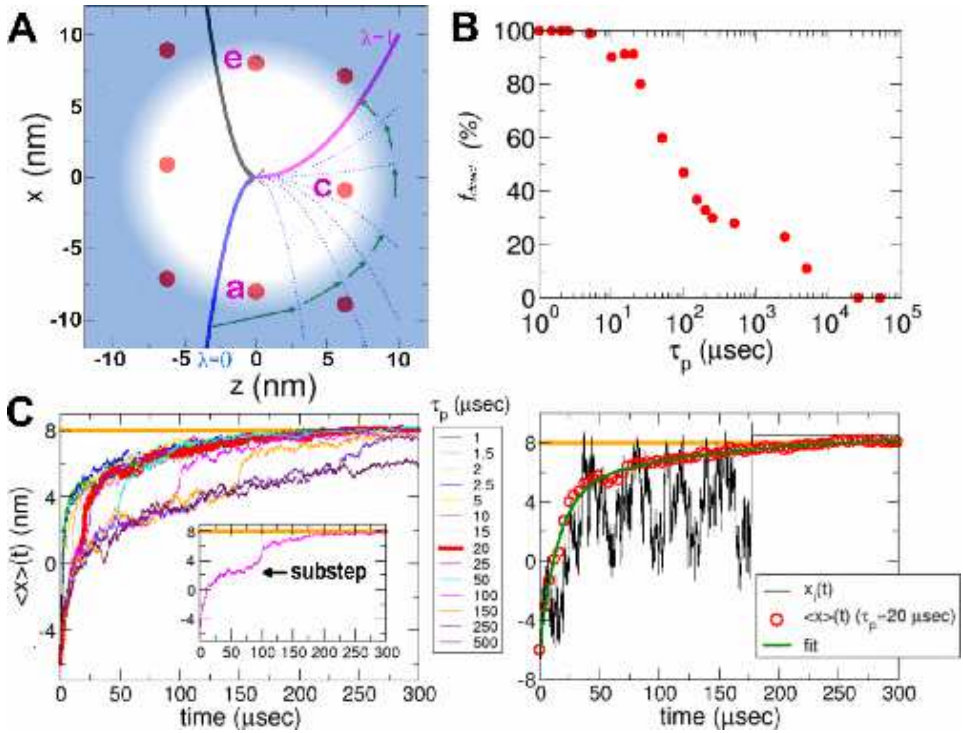}
\caption{Results of the Brownian dynamics simulations of quasi-particle representing the tethered head on the PMF define in Eq.\ref{eqn:tPMF}.
{\bf A.} Conceptual representation of kinesin's stepping dynamics (see text).
{\bf B.} The fraction of trajectories ($f_{direct}$) directly reaching the target binding site $e$ as a function of $\tau_{p}$.
{\bf C.} (Left) The ensemble average of 100 trajectories generated for varying $\tau_{p}$. Substeps are manifested for $\langle x\rangle(t)$s with $\tau_p>20$ $\mu s$. The inset shows the substep in the $\langle x\rangle(t)$ for $\tau_p=100$ $\mu s$.
(Right) An actual time trace $x_i(t)$ is plotted in black with the time interval of 0.1 $\mu s$ from which the ensemble average is obtained as $\langle x\rangle(t)=\sum_{i=1}^{100}x_i(t)$.
The ensemble average of trajectories for $\tau_{p}=20.0$ $\mu s$ is fitted to $\langle x\rangle(t)=14.7$ nm$\times [1-0.74e^{-t/15.3\mu s}-0.26e^{-t/149\mu s}]-6.0$ nm.
\label{brownian}}
\end{figure}
\begin{figure}[ht]
\includegraphics[width=5.00in]{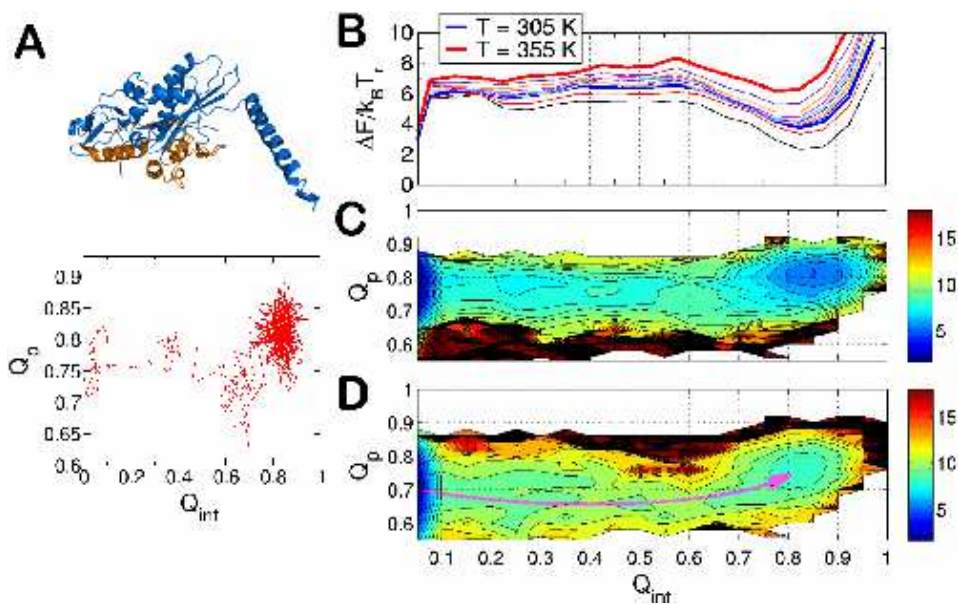}
\caption{PMF between the tethered head and the MT-binding site $e$ as a function of two order parameters, $Q_p$ and $Q_{int}$.
{\bf A.} (Top) The MT-binding motifs of the kinesin are colored in orange. (Bottom) An exemplary binding trajectory as functions of $Q_p$ and $Q_{int}$.
{\bf B.} $\Delta F(Q_{int})$, 1-D free energy profile as a function of $Q_{int}$ at varying temperatures.
{\bf C.} $\Delta F(Q_{int},Q_p)$ at $T=305$ K$(=T_r)$. As the binding progresses, minor structural disruption is observed. The free energy difference is color-coded from blue to red in $k_BT_r$ unit.
{\bf D.} $\Delta F(Q_{int},Q_p)$ at $T=355$ K. The partial unfolding along the binding process is more pronounced than at a lower temperature.
The overall binding pathway is drawn with a curved arrow.
\label{Q_Qint_map}}
\end{figure}

\renewcommand{\thefigure}{7}
\begin{figure}[ht]
\includegraphics[width=2.50in]{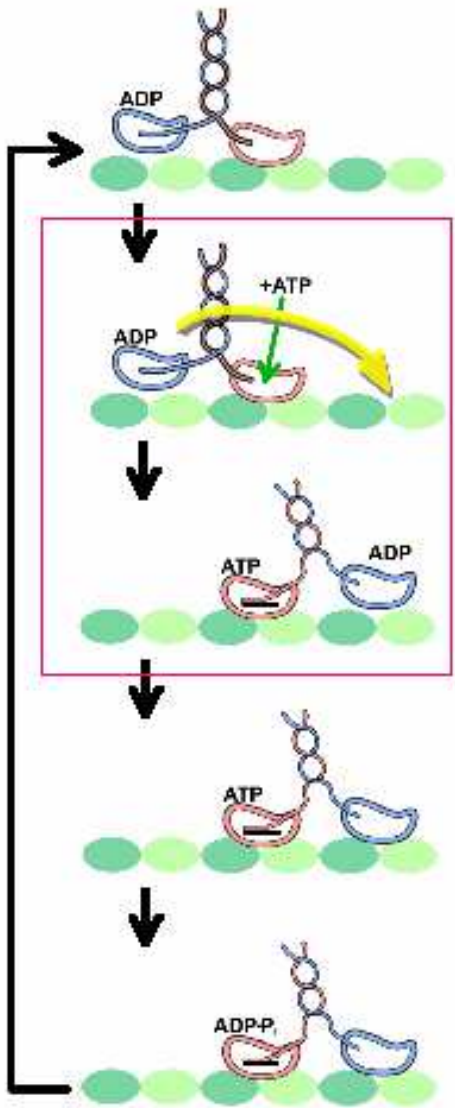}
\caption{Mechanochemical cycle of the conventional kinesin at the track of MT protofilament. 
ATP binding to the leading head (red) induces a undocked-to-docked transition of the neck-linker, which results in 16-nm stepping dynamics of the trailing head (blue). 
When the tethered head succeeds in binding to the next binding site, the rearward tension built on the neck-linker perturbs the nucleotide binding 
site and eases the dissociation of ADP from the tethered head \cite{BlockPNAS06,Hyeon07PNAS}. The ATP hydrolysis follows at the trailing head (red). 
The step corresponding to the stepping motion is enclosed by a box. 
\label{cycle}}
\end{figure}
\renewcommand{\thefigure}{8}
\begin{figure}[ht]
\includegraphics[width=7.00in]{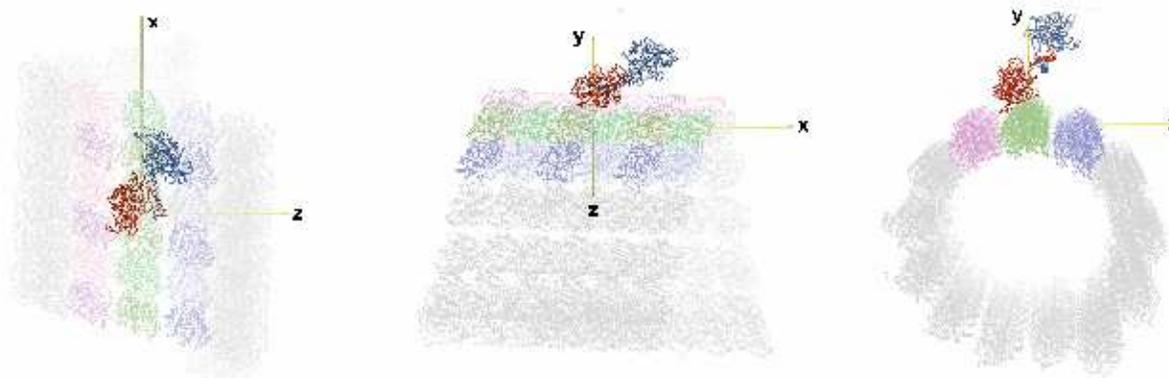}
\caption{A snapshot of simulation shown in the xz, xy, yz plane. The kinesin 
monomer in red is the MT-bound head, and the kinesin monomer in blue is the tethered head. 
\label{MT_and_kinesin_SI}}
\end{figure}
\renewcommand{\thefigure}{9}
\begin{figure}[ht]
\includegraphics[width=5.00in]{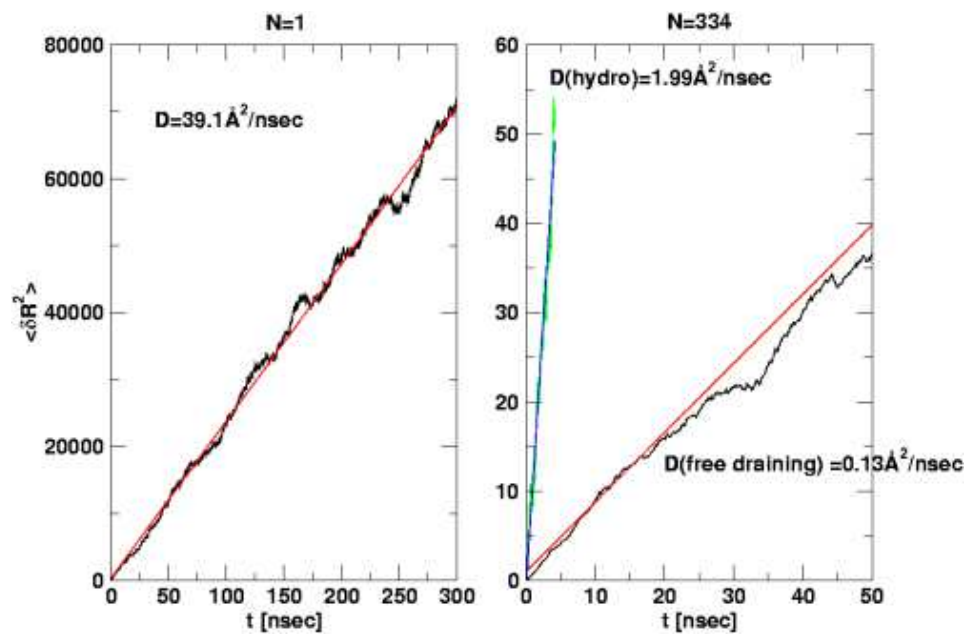}
\caption{The diffusion constant of kinesin head domain is computed using Brownian dynamics simulations with explicit hydrodynamic tensor (Rotne-Prager tensor). Mean square displacements ($\langle\delta R^2\rangle$) of single bead (left) and kinesin head composed of 334 beads (right) are compared. For kinesin head, we performed Brownian dynamics simulations with and without hydrodynamics. By including the hydrodynamics, the correct diffusion behavior of kinesin head is obtained. \label{hydro}}
\end{figure}
\renewcommand{\thefigure}{10}
\begin{figure}[ht]
\includegraphics[width=5.00in]{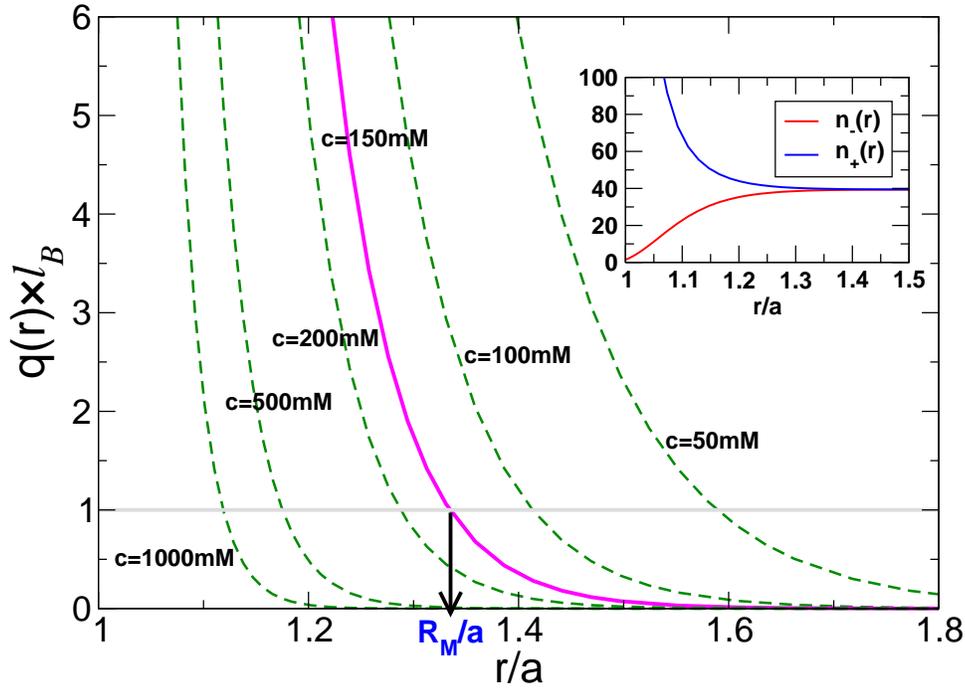}
\caption{The numerical solution of nonlinear Poisson-Boltzmann equation under varying concentrations of monovalent salt. The electrostatic potential around cylinderical geometry $\phi(r)$ is obtained by solving $\nabla^2\phi=\kappa^2\sinh{\phi}$ with boundary condition, and $\phi(r)$ is used to compute $n_+(r)$, $n_-(r)$, and $q(r)$. 
The inset and the purple line are the numerical solution for $c=150$ mM. 
The thickness of condense ion layer, $R_M$, is determined using the Manning condensation criterion, $q(R_M)\times l_B=1$. \label{condensation}}
\end{figure}


\begin{thebibliography}{10}

\bibitem{Brady85Nature}
Brady, S.~T. (1985) {\em Nature} {\bf 317}, 73--75.

\bibitem{Vale85Cell}
Vale, R.~D., Reese, T.~S., \& Sheetz, M.~P. (1985) {\em Cell} {\bf 42}, 39--50.

\bibitem{Visscher99Science}
Visscher, K., Schnitzer, M.~J., \& Block, S.~M. (1999) {\em Nature} {\bf 400},
  184--187.

\bibitem{Block03Science}
Asbury, C.~L., Fehr, A.~N., \& Block, S.~M. (2003) {\em Science} {\bf 302},
  2130--2134.

\bibitem{Cross05Nature}
Carter, N.~J. \& Cross, R.~A. (2005) {\em Nature} {\bf 435}, 308--312.

\bibitem{BlockPNAS06}
Guydosh, N.~R. \& Block, S.~M. (2006) {\em Proc. Natl. Acad. Sci.} {\bf 103},
  8054--8059.

\bibitem{Kaseda03NCB}
Kaseda, K., Higuchi, H., \& Hirose, K (2003) {\em Nature Cell Biol.} {\bf 5},
  1079--1082.

\bibitem{Yildiz04Science}
Yildiz, A., Tomishige, M., Vale, R.~D., \& Selvin, P.~R. (2004) {\em Science}
  {\bf 303}, 676--678.

\bibitem{Block03PNAS}
Block, S.~M., Asbury, C.~L., Shaevitz, J.~W., \& Lang, M.~J. (2003) {\em Proc.
  Natl. Acad. Sci.} {\bf 100}, 2351--2356.

\bibitem{Gilbert94Biochem}
Gilbert, S.~P. \& Johnson, K.~A. (1994) {\em Biochemistry} {\bf 33},
  1951--1960.

\bibitem{Taylor95Biochem}
Ma, Y.~Z. \& Taylor, E.~W. (1995) {\em Biochemistry} {\bf 34}, 13242--13251.

\bibitem{Moyer98BC}
Moyer, M.~L., Gilbert, S.~P., \& Johnson, K.~A. (1998) {\em Biochemistry} {\bf
  37}, 800--813.

\bibitem{Cross04TBS}
Cross, R.~A. (2004) {\em Trends Biochem. Sci.} {\bf 29}, 301--309.

\bibitem{Prost97RMP}
J{\"u}licher, F., Ajdari, A., \& Prost, J. (1997) {\em Rev. Mod. Phys.} {\bf
  69}, 1269--1281.

\bibitem{Fisher01PNAS}
Fisher, M.~E. \& Kolomeisky, A.~B. (2001) {\em Proc. Natl. Acad. Sci.} {\bf
  98}, 7748--7753.

\bibitem{Fisher05PNAS}
Fisher, M.~E. \& Kim, Y.~C. (2005) {\em Proc. Natl. Acad. Sci.} {\bf 102},
  16209--16214.

\bibitem{Gao06PNAS}
Shao, Q. \& Gao, Y.~Q. (2006) {\em Proc. Natl. Acad. Sci.} {\bf 103},
  8072--8077.

\bibitem{Kolomeisky07ARPC}
Kolomeisky, A.~B. \& Fisher, M.~E. (2007) {\em Ann. Rev. Phys. Chem.} {\bf 58},
  675--695.

\bibitem{Hyeon07PNAS}
Hyeon, C. \& Onuchic, J.~N. (2007) {\em Proc. Natl. Acad. Sci.} {\bf 104},
  2175--2180.

\bibitem{Block07BJ}
Block, S.~M. (2007) {\em Biophys. J.} {\bf 92}, 2986--2995.

\bibitem{Nishiyama01NCB}
Nishiyama, M., Muto, E., Inoue, Y., Yanagida, T., \& Higuchi, H. (2001) {\em
  Nature Cell Biol.} {\bf 3}, 425--428.

\bibitem{Coppin96PNAS}
Coppin, C.~M., Finer, J.~T., Spudich, J.~A., \& Vale, R.~D. (1996) {\em Proc.
  Natl. Acad. Sci.} {\bf 93}, 1913--1917.

\bibitem{Howard93JCB}
Ray, S., Meyhofer, E., Milligan, R.~A., \& Howard, J. (1993) {\em J. Cell
  Biol.} {\bf 121}, 1083--1093.

\bibitem{Sheetz95BJ}
Wang, Z., Khan, S., \& Sheetz, M.~P. (1995) {\em Biophys. J.} {\bf 69},
  2011--2023.

\bibitem{Hancock99PNAS}
Hancock, W.~D. \& Howard, J. (1999) {\em Proc. Natl. Acad. Sci.} {\bf 96},
  13147--13152.

\bibitem{ValeNature99}
Rice, S., Lin, A.~W., Safer, D., Hart, C.~L., Naber, N., Carragher, B.~O.,
  Cain, S.~M., Pechatnikova, E., {Wilson-Kubalek}, E.~M., Whittaker, M., Pate,
  E., Cooke, R., Taylor, E.~M., Milligan, R.~A., \& Vale, R.~D. (1999) {\em
  Nature} {\bf 402}, 778--784.

\bibitem{Asenjo06NSMB}
Asenjo, A.~B., Weinberg, Y., \& Sosa, H. (2006) {\em Nature Struct. Mol. Biol.}
  {\bf 13}, 648--654.

\bibitem{Tomishige06NSMB}
Tomishige, M., Stuurman, N., \& Vale, R.~D. (2006) {\em Nature Struct. Mol.
  Biol.} {\bf 13}, 887--893.

\bibitem{Downing02Structure}
Le, H., {deRosier}, D., Nicholson, W., Nogales, E., \& Downing, K. (2002) {\em
  Structure} {\bf 10}, 1317--1328.

\bibitem{Best05Structure}
R.~B.~Best, Y.~G.~Chen \& Hummer, G. (2005) {\em Structure} {\bf 13},
  1755--1763.

\bibitem{Hyeon06PNAS}
Hyeon, C., Lorimer, G.~H., \& Thirumalai, D. (2006) {\em Proc. Natl. Acad.
  Sci.} {\bf 103}, 18939--18944.

\bibitem{Miyashita03PNAS}
Miyashita, O., Onuchic, J.~N., \& Wolynes, P.~G. (2003) {\em Proc. Natl. Acad.
  Sci.} {\bf 100}, 12570--12575.

\bibitem{Shoemaker00PNAS}
Shoemaker, B.~A., Portman, J.~J., \& Wolynes, P.~G. (2000) {\em Proc. Natl.
  Acad. Sci.} {\bf 97}, 8868--8873.

\bibitem{Dyson05NRMCB}
Dyson, H.~J. \& Wright, P.~E. (2005) {\em Nature Rev. Mol. Cell Biol.} {\bf 6},
  197--208.

\bibitem{Levy07JACS}
Levy, Y., Onuchic, J.~N., \& Wolynes, P.~G. (2007) {\em J. Am. Chem. Soc.} {\bf
  129}, 738--739.

\bibitem{Peskin93BJ}
Peskin, C.~S., Odell, G.~M., \& Oster, G.~F. (1993) {\em Biophys. J.} {\bf 65},
  316--324.

\bibitem{Yanagida05NatureChemBiol}
Taniguchi, Y., Nishiyama, M., Yoshiharu, I., \& Yanagida, T. (2005) {\em
  Nature. Chem. Biol.} {\bf 1}, 342--347.

\bibitem{ThirumJPI}
Thirumalai, D. (1995) {\em J. Phys. I (Fr.)} {\bf 5}, 1457--1467.

\bibitem{OnuchicCOSB04}
Onuchic, J.~N. \& Wolynes, P.~G. (2004) {\em Curr. Opin. Struct. Biol.} {\bf
  14}, 70--75.

\bibitem{Eaton04COSB}
Kubelka, J., Hofrichter, J., \& Eaton, W.~A. (2004) {\em Curr. Opin. Struct.
  Biol.} {\bf 14}, 76--88.

\bibitem{LI04Polymer}
Li, M.~S., Klimov, D.~K., \& Thirumalai, D. (2004) {\em Polymer} {\bf 45},
  573--579.

\bibitem{Bryngelson90Biopolymers}
Bryngelson, J.~D. \& Wolynes, P.~G. (1990) {\em Biopolymers} {\bf 30},
  177--188.

\bibitem{Uhlenbeck30PR}
Uhlenbeck, G.~E. \& Ornstein, L.~S. (1930) {\em Phys. Rev.} {\bf 36}, 823--841.

\bibitem{HyeonBC05}
Thirumalai, D. \& Hyeon, C. (2005) {\em Biochemistry} {\bf 44}(13), 4957--4970.

\bibitem{Zwanzig90ACR}
Zwanzig, R. (1990) {\em Acc. Chem. Res.} {\bf 23}, 148--152.

\bibitem{Hyeon06Structure}
Hyeon, C., Dima, R.~I., \& Thirumalai, D. (2006) {\em Structure} {\bf 14},
  1633--1645.

\bibitem{Manning69JCP}
Manning, G.~S. (1969) {\em J. Chem. Phys.} {\bf 51}(3), 924--933.

\bibitem{BishopJCP91}
Bishop, M., Clarke, J.~H.~R., Rey, A., \& Freire, J.~J. (1991) {\em J. Chem.
  Phys.} {\bf 95}, 4589--4592.

\bibitem{Hirokawa00PNAS}
Okada, Y. \& Hirokawa, N. (2000) {\em Proc. Natl. Acad. Sci.} {\bf 97},
  640--645.

\bibitem{HirokawaSCI04}
Nitta, R., Kikkawa, M., Okada, Y., \& Hirokawa, N. (2004) {\em Science} {\bf
  305}, 678--683.

\bibitem{SwendsenPRL88}
Ferrenberg, A.~M. \& Swendsen, R.~H. (1988) {\em Phys. Rev. Lett.} {\bf
  61}(23), 2635--2638.

\bibitem{KumarJCC1992}
Kumar, S., Bouzida, D., Swendsen, R.~H., Kollman, P.~A., \& Rosenberg, J.~M.
  (1992) {\em J. Comp. Chem.} {\bf 13}(8), 1011--1021.

\bibitem{KremerJCP93}
D{\"u}nweg, B. \& Kremer, K. (1993) {\em J. Chem. Phys.} {\bf 99}, 6983--6997.

\bibitem{Rotne69JCP}
Rotne, J. \& Prager, S. (1969) {\em J. Chem. Phys.} {\bf 50}, 4831--4837.

\bibitem{Manning69JCP}
Manning, G.~S. (1969) {\em J. Chem. Phys.} {\bf 51}(3), 924--933.

\bibitem{Gilbert95Nature}
Gilbert, S.~P., Webb, M.~R., Brune, M., \& Johnson, K.~A. (1995) {\em Nature}
  {\bf 373}, 671.

\bibitem{BlockPNAS06}
Guydosh, N.~R. \& Block, S.~M. (2006) {\em Proc. Natl. Acad. Sci.} {\bf 103},
  8054--8059.

\end{thebibliography}
\end{document}